\documentclass[journal]{IEEEtran}

\usepackage{amssymb}
\usepackage{mathtools}
\usepackage{diagbox}
\usepackage{graphicx,cite,epsfig,amssymb,amsmath}
\usepackage{pifont}
\usepackage{bm}
\usepackage{color}
\usepackage{stmaryrd}        
\usepackage{standalone}
\usepackage{preview}           
\usepackage{xcolor}
\usepackage{bbding}

\usepackage{booktabs}
\usepackage{listings}
\usepackage{multirow}
\usepackage[colorlinks,linkcolor=black,anchorcolor=black,citecolor=black]{hyperref}

\usepackage[makeroom]{cancel} 
\usepackage{amsthm}
\usepackage{bbm} 
\usepackage[caption=false, font=footnotesize]{subfig} 
\usepackage{adjustbox}

\usepackage{titlesec}

\usepackage{hyperref}

\UseRawInputEncoding

\begin{document}
\title{A Monte-Carlo Based Construction of Polarization-Adjusted Convolutional (PAC) Codes}

\author{Mohsen~Moradi\textsuperscript{\href{https://orcid.org/0000-0001-7026-0682}{\includegraphics[scale=0.06]{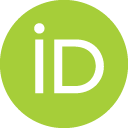}}},
Amir~Mozammel\textsuperscript{\href{https://orcid.org/0000-0003-3474-9530}{ \includegraphics[scale=0.06]{figs/ORCID}}}
\thanks{The authors are with the Department of Electrical-Electronics Engineering, Bilkent University, Ankara TR-06800, Turkey (e-mail: moradi@ee.bilkent.edu.tr, a.mozammel@ee.bilkent.edu.tr).}%
}

\maketitle
\begin{abstract}
This paper proposes a rate-profile construction method for polarization-adjusted convolutional (PAC) codes of any code length and rate, which is capable of maintaining trade-off between the error-correction performance and decoding complexity of PAC code.
The proposed method can improve the error-correction performance of PAC codes while guaranteeing a low mean sequential decoding complexity for signal-to-noise ratio (SNR) values beyond a target SNR value.
\end{abstract}


\section{Introduction}

\IEEEPARstart{P}{olarization-adjusted} convolutional (PAC) codes are a new class of linear codes that combine polar and convolutional coding \cite{arikan2019sequential}.
Benefiting from channel polarization and convolutional coding, PAC codes outperform both polar codes and convolutional codes (CCs) for short block lengths and certain code rates \cite{moradi2020PAC}. 

The first step of encoding of PAC codes is to insert a source data of length $K$ into a data carrier word of length $N$ in accordance with a data index set $\mathcal{A}$.
The data index set $\mathcal{A}$ is also known as the rate profile of PAC codes with a code rate of $R = K/N$.
We refer to the selection of $K$ indices out of $N$ possible indices as the rate-profile construction.
The rate-profile construction plays an important role on error-correction performance and computational complexity of PAC codes.
Two well known rate profiles for PAC codes are polar and Reed-Muller (RM) rate profiles \cite{arikan2019sequential}.
Under sequential decoding, when using polar rate profile, PAC codes have low computational complexity, but suffer from poor error-correction performance, 
whereas using RM rate profile, improves the error-correction performance of PAC codes, but increases its computational complexity \cite{arikan2019sequential}.

Sequential decoding \cite{wozencraft1957sequential} is a depth-first tree search algorithm.
Two well known sequential algorithms are Fano \cite{fano1963heuristic} and stack algorithms \cite{zigangirov1966some}, \cite{jelinek1969fast}.
During a decoding session, stack algorithm visits each node at most once, but requires relatively large storage capacity, whereas the Fano algorithm may have repeated visits for nodes, but requires lower memory. 
Due to its low memory requirement, the Fano decoding is more favorable for practical hardware implementations of PAC codes \cite{mozammel2020hardware}.
Since both algorithms eventually choose the same paths on the decoding tree, the set of nodes visited by Fano and stack algorithms are identical \cite{geist1970algorithmic}. 
In this paper, we use Fano algorithm in our simulations.

One drawback of sequential decoding is its variable search complexity which heavily depends on the channel cutoff rate $R_0$. 
For CC codes with code rates above channel cutoff rate ($R > R_0$), the search complexity of sequential decoder increases exponentially, whereas for $R < R_0$, sequential decoder has a finite search complexity.
Similar to the sequential decoding of conventional CCs, sequential decoding of PAC codes has a variable computational complexity and suffers from a cutoff rate phenomenon \cite{moradi2020PAC}

Taking advantage of cutoff rate phenomenon, we introduce a Monte-Carlo based rate-profile construction method for PAC codes for which the mean computational complexity of sequential decoder is guaranteed to be finite for signal-to-noise ratio (SNR) values above a target SNR.
The simulation results confirm this statement and show an error-correction performance improvement for PAC codes constructed with our proposed method.

In this paper, vectors are denoted by boldface letters.
We use $\mathbf{u}^i$ to indicate subvector $(u_1, \cdots, u_i)$.
For any subset $\mathcal{A}$ of indices  $\{1, 2, ..., N\}$ its cardinality is denoted by $|\mathcal{A}|$ and its complement is expressed by $\mathcal{A}^c$.
The subvector $\mathbf{u}_{\mathcal{A}}$ represents $(u_i : i\in \mathcal{A})$.
We define a 1-bit quantization function $q(x,\delta)$ as
\begin{equation*}
    q(x,\delta) :=
    \begin{cases}
    1, &\text{if}~x\geq \delta,\\
    0, &\text{otherwise},
    \end{cases}
\end{equation*}
where $0 < \delta < 1$.

The remainder of this paper is organized as follows.
Section \ref{sec: background} briefly reviews the channel polarization phenomenon.
Section \ref{sec: scheme} gives an overview of PAC coding scheme.
In section \ref{sec: construction}, a rate-profile construction method for PAC codes is introduced. 
Section \ref{sec: simulation} provides simulation results.
Finally, section \ref{sec: conclusion} concludes this paper.


\section{Channel Polarization} \label{sec: background}
Channel polarization is the operation of receiving $N$ independent copies of a binary-input discrete memoryless channel (B-DMC) and producing $N$ new channels that show polarization effect such that the individual capacities of new (polarized) channels approach either 0 or 1, asymptotically \cite{arikan2009channel}.
Let $W: \mathcal{X} \longrightarrow \mathcal{Y}$ denote a B-DMC with an arbitrary output alphabet $\mathcal{Y}$. 
The channel transition probability is defined by $W(y|x)$, where $x \in \mathcal{X}$ and $y \in \mathcal{Y}$.
The polarized bit-channels $W_N^{(i)}$ are defined as
$W_N^{(i)} : \mathcal{X} \longrightarrow \mathcal{Y} \times \mathcal{X}^{i-1}$ for $1 \leq i \leq N$.

One important parameter of bit-channel $W_N^{(i)}$ is the Bhattacharyya parameter which is defined as
\begin{equation}
\begin{split}
    &Z(W_N^{(i)}) =\\
    & \sum_{\mathbf{y} \in \mathcal{Y}^N}\sum_{\mathbf{u}^{i-1} \in \mathcal{X}^{i-1}} \sqrt{W_N^{(i)}(\mathbf{y},\mathbf{u}^{i-1},|0)W(\mathbf{y},\mathbf{u}^{i-1}|1)},
\end{split}
\end{equation}
from which the bit-channel cutoff rate $E_0(1, W_N^{(i)})$ can be obtained by
\begin{equation}
    E_0(1, W_N^{(i)}) = 1 - \log_2{}\left( 1 + Z(W_N^{(i)}) \right).
\end{equation}
For AWGN channel, bit-channel Bhattacharyya parameters can be obtained using methods such as Gaussian approximation \cite{li2013practical}.

\begin{figure}[t] 
\centering
	\includegraphics [width = \columnwidth]{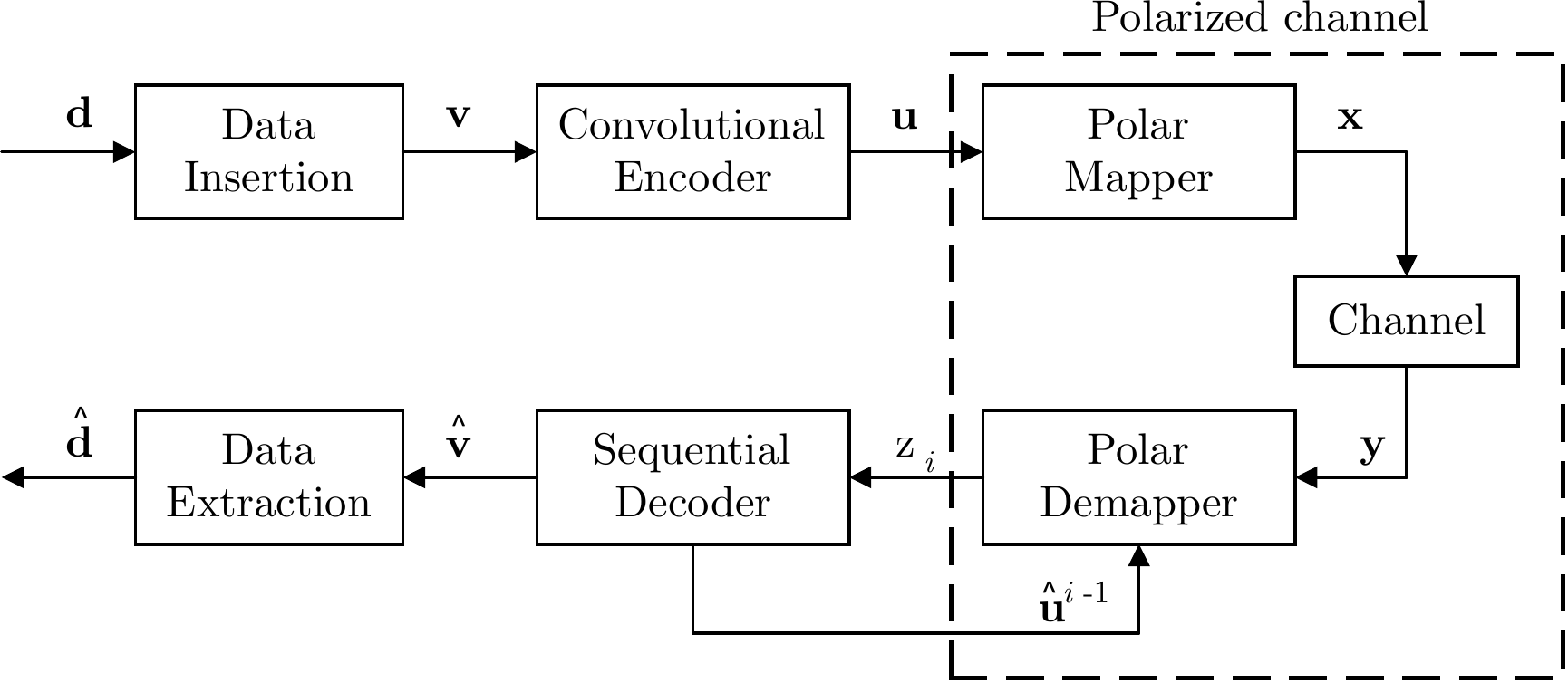}
	\caption{Block diagram of PAC coding scheme.} 
	\label{fig: flowchart}
\end{figure}

\section{PAC Coding Scheme} \label{sec: scheme}
Fig. \ref{fig: flowchart} shows a block diagram of PAC coding scheme.
For a $(N,K,\mathcal{A}, \mathbf{T})$ PAC code, the parameters $N$ and $K$ are codeword and message lengths, respectively, $\mathcal{A}$ is the rate profile, and $\mathbf{T}$ is an upper-triangular Toeplitz matrix constructed with a connection polynomial $\mathbf{g}$.
$\mathbf{d}$ is the message generated uniformly at random over all possible source data of length $K$ in a binary field $\mathbb{F}_2$.
The data insertion block maps these $K$ bits into a data carrier vector $\mathbf{v}$ in accordance with the data set $\mathcal{A}$, thus inducing a code rate of $R=K/N$.
More precisely, $\mathbf{v}$ is obtained by $\mathbf{v}_\mathcal{A} = \mathbf{d}$ and $\mathbf{v}_{\mathcal{A}^c} = 0$.
The bits that are fixed to zero are called frozen bits, and the rest are called information bits. 
Then $\mathbf{v}$ is sent to the convolutional encoder and encoded as $\mathbf{u} = \mathbf{v}\mathbf{T}$.
Then $\mathbf{u}$ is transformed into $\mathbf{x}$ with standard polar transformation $\mathbf{x}=\mathbf{u}\mathbf{F}^{\otimes n}$, where $\textbf{F}^{\otimes n}$ is the $n$th kroneker power of $\textbf{F} = \begin{bsmallmatrix} 1 & 0\\ 1 & 1 \end{bsmallmatrix}$ with $n = \log_2 N$.
After polar transformation, $\mathbf{x}$ is sent through the channel.
Polar demapper receives channel output $\mathbf{y}$ and previously decoded bits and calculates the log-likelihood ratio (LLR) value of current bit $z_i$.
Finally, the sequential decoder outputs an estimate of carrier word $\hat{\mathbf{v}}$, from which the $K$-bits data can be extracted according to $\mathcal{A}$ \cite{arikan2019sequential}. 


\section{Monte Carlo Construction of PAC Codes}\label{sec: construction}
The mean computational complexity of sequential decoding of CCs is finite for the code rates below the channel cutoff rate (i.e. $R < E_0(1,W)$) \cite[p.~279]{gallager1968information}.
Note that the CC of PAC codes sees a polarized channel with bit-channel cutoff rate $E_0(1, W_N^{(i)})$ and has a time-varying code rate.
For this reason, the condition of having finite mean computational complexity of CCs may be extended to PAC codes as
\begin{equation}\label{eq: condition}
    l R_l < \sum_{i=1}^{l} E_0(1,W_N^{(i)})
\end{equation}
for all $1 \leq l \leq N$, where $R_l = \frac{\lambda_l}{l}$ with $\lambda_l$ being the number of information bits in the first $l$ bits of $\mathbf{v}$ \cite{moradi2020metric}.

\begin{figure}[t] 
\centering
	\includegraphics [width = \columnwidth]{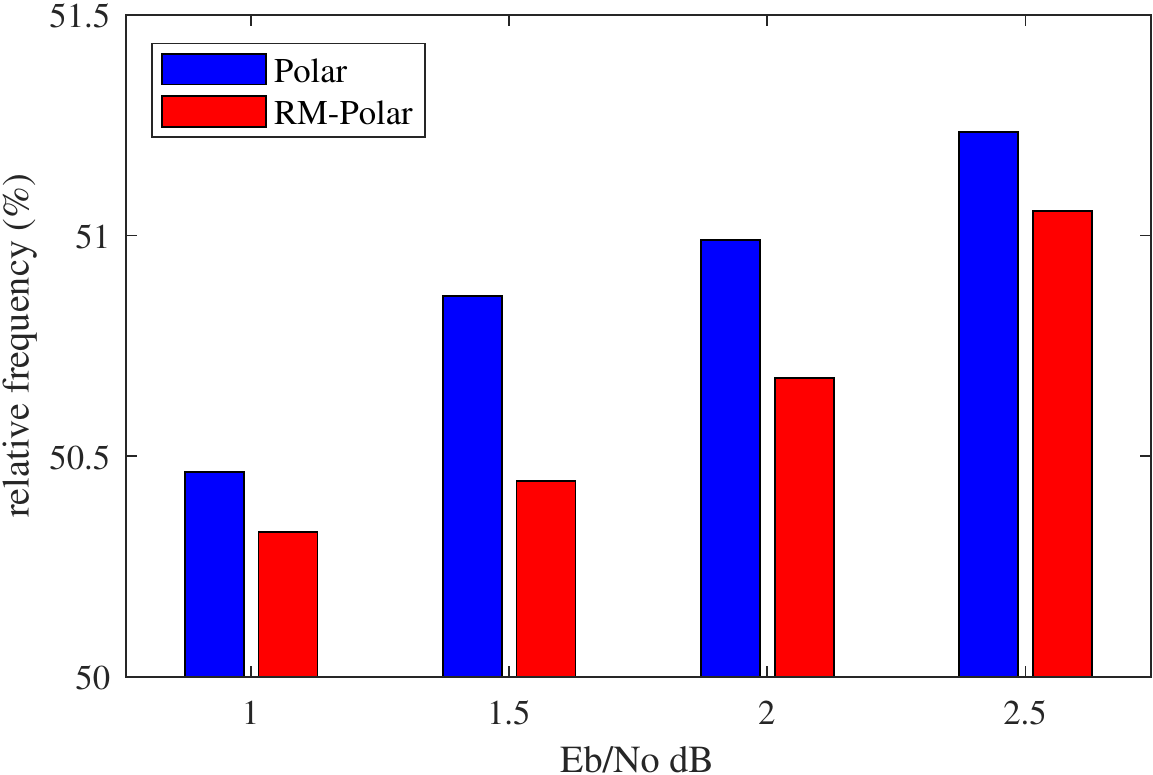}
	\caption{Error frequency of subsequent bits of FBE for PAC$(256, 128)$ code.} 
	\label{fig: distribution}
\end{figure}

Due to the existence of memory in CCs, for a CC of memory size $m$ and rate $1/b$, each bit of message can affect up to $b(m+1)$ bits of output vector and an error in a bit may introduce burst errors of length up to $b(m+1)$ \cite[p.~416]{wozencraft1957sequential}.
Similarly, due to the existence of polarized channel with memory of $N$, the occurrence of burst errors up to length $N$ are possible for PAC codes.
For this reason, if the first bit error (FBE) occurs at index $i$, the following bits from $i+1$ to $N$ will be decoded wrongly with a high probability.
Fig. \ref{fig: distribution} shows the average fraction of the subsequent bits after FBE that are decoded wrongly for a PAC$(256, 128)$ code.
These values are extracted from $10^{4}$ decoding failures of PAC$(256, 128)$ code with polar and RM-Polar rate profiles.
The polar rate profile is obtained by choosing the $K$ most reliable indices evaluated at $E_b/N_0 = 2.5$ dB \cite{arikan2009channel}.
To form the RM-Polar rate profile, first, all 93 row indices of matrix $\textbf{F}^{\otimes 8}$ with Hamming weights more than or equal to 32 are selected. 
Then, out of the 70 rows of $\textbf{F}^{\otimes 8}$ with weights equal to 16, 35 row indices with the highest reliabilities are selected \cite{li2014rm}. 
As this figure illustrates, at all the simulated SNR values, more than half of the subsequent bits of FBE are decoded wrongly for both of the rate profiles.

Considering the condition \eqref{eq: condition} and the results of Fig. \ref{fig: distribution}, it is possible to construct a rate profile $\mathcal{A}$ that results in a relatively good error-correction performance while guaranteeing low mean decoding complexity for SNR values beyond a desired SNR value.
To achieve this goal, we propose the following Monte Carlo (MC) procedure for a PAC code of length $N$, code rate $R$, and quantization level $\delta$.

\begin{table*}[ht] \label{tab: rate profile}
\centering
\caption{Rate profiles.}
\renewcommand{\arraystretch}{1.2}
\begin{tabular}{ccc}
\hline
$(N,K)$                      & $E_b/N_0$ (dB) &  Rate profile $\boldsymbol{\alpha}$ (hexadecimal)                                                 \\ \hline
\multirow{3}{*}{$(256,128)$} & 1.5      & 00000000000001170001013F037F7FFF0001017F077F7FFF177F7FFF7FFFFFFF \\ \cline{2-3} 
                           & 2.5      & 000000010001011F0001013F077FFFFF0001037F177F7FFF011F1FFF7FFFFFFF \\ \cline{2-3} 
                           & 3        & 000000010001013F0001037F077FFFFF0001077F177F7FFF013F1FFF177F7FFF \\ \hline
\multirow{2}{*}{$(64,32)$}   & 3        & 0001017F017F7FFF                                                 \\ \cline{2-3} 
                           & 5        & 0007077F031F17FF                                                 \\ \hline
\end{tabular}
\end{table*}

\begin{enumerate}
    \item Choose a target SNR value which satisfies
\begin{equation*}
    NR < \sum_{i=1}^{N} q(E_0(1,W_N^{(i)}),\delta).
\end{equation*}
    
    \item Construct a rate profile $\mathcal{A}$ such that $i \in \mathcal{A}$ if $q(E_0(1,W_N^{(i)}),\delta) = 1$ for $i = 1,\cdots,N$.
    \item Construct an all-zero vector $\mathbf{h} = (h_1, \cdots, h_N)$ to store FBE index frequencies.
    \item Run Monte Carlo simulation for a PAC$(N,|\mathcal{A}|,\mathcal{A},\mathbf{T})$ code.
    At the end of each iteration of Monte Carlo simulation, find $j$ such that $j = \min(i): v_i \neq \hat{v}_i$ and update $\mathbf{h}$ by $h_j \leftarrow h_j + 1$.
    \item Find the index $j$ of maximum element of $\mathbf{h}$ and remove $j$ from $\mathcal{A}$.
    \item If $|\mathcal{A}| > K$, go to step 3.
    \item Return $\mathcal{A}$.
\end{enumerate}

Step 1 follows \eqref{eq: condition} and guarantees a low mean computational complexity for the sequential decoder of the PAC code for the SNR values beyond the target SNR.
Step 4 is inspired from the results of Fig. \ref{fig: distribution} and aims to obtain a good error-correction performance by avoiding the assignment of message to those indices of data carrier vector $\mathbf{v}$ that are highly probable to be FBE (hence, reduces the burst error occurrence frequency).

 \begin{figure}[t] 
\centering
	\includegraphics [width = \columnwidth]{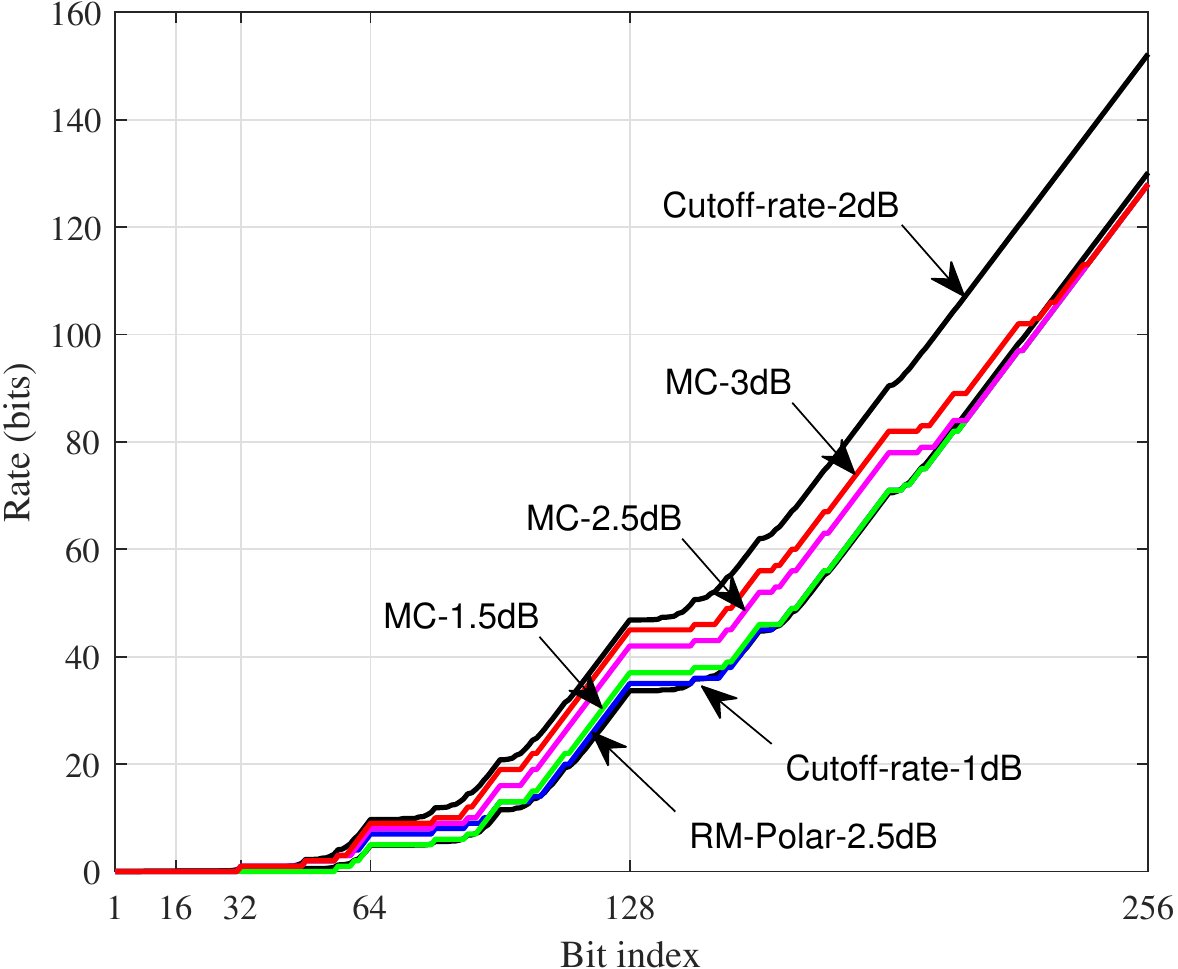}
	\caption{PAC$(256, 128)$ code rate profiles.} 
	\label{fig: profile_N256}
\end{figure}

\begin{figure}[t] 
\centering
	\includegraphics [width = \columnwidth]{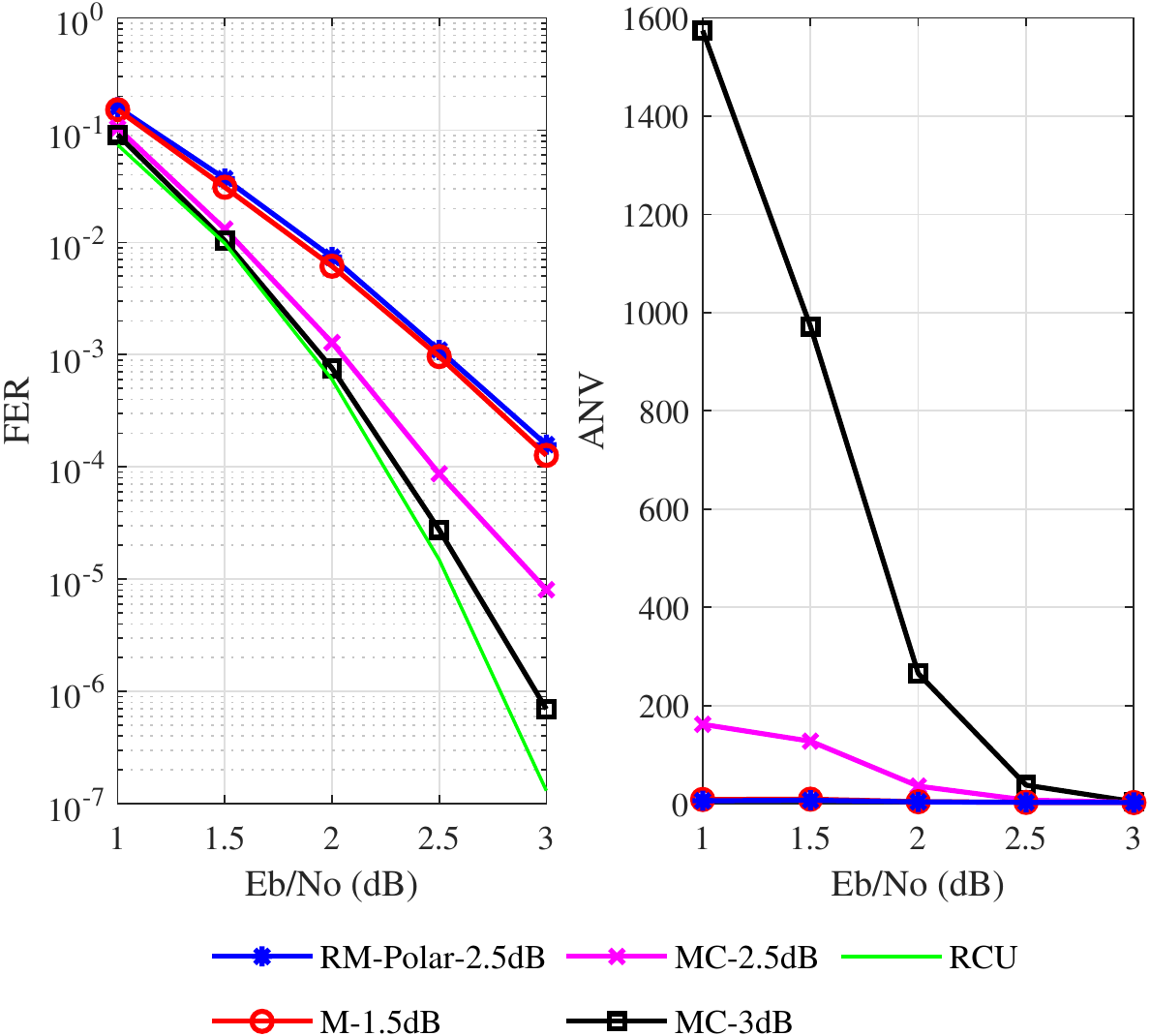}
	\caption{Error-correction performance and complexity of PAC$(256, 128)$ codes.} 
	\label{fig: FER_ANV_N256}
\end{figure}

\section{Simulation results} \label{sec: simulation}
This section provides simulation results for PAC$(256, 128)$ and PAC$(64, 32)$ codes with the rate profiles obtained by our proposed method.
These results are obtained using Fano decoder with connection polynomial $\mathbf{g} = 3211$, threshold spacing $\Delta = 2$, and metric function of \cite{moradi2020metric} over binary-input additive white Gaussian noise channel.

We compare the decoding complexity of Fano decoder with different rate profiles by a random variable $\Theta$, which counts the number of times each bit is visited by the decoder during a decoding session \cite[p.~444]{wozencraft1957sequential}.
We are primarily interested in the expectation of $\Theta$ per bit (i.e. $\mathbb{E}(\Theta)/N$), which we call average number of visits (ANV) and estimate it by computing the empirical mean of $\Theta$ over a sufficiently large number of simulation runs and dividing the result by code length $N$.

For PAC$(256, 128)$ code, three different rate profiles are obtained using our proposed construction method with $\delta = 0.5$ and $E_b/N_0 = 1.5,2.5$ and $3$ dB.
For these SNR values, at step 2 of algorithm, $|\mathcal{A}|$ is $144, 165$ and $176$, and to obtain the final rate profile the algorithm runs for $16, 37$ and $48$ iterations, respectively. 
Table \ref{tab: rate profile} provides the obtained rate profiles.
For compact representation of the rate profiles, we use a binary vector $\boldsymbol{\alpha}$ of length $N$ such that $\alpha_i = 1$ if $i \in \mathcal{A}$ and $\alpha_i = 0$, otherwise, for $1\leq i \leq N$.

Fig. \ref{fig: profile_N256} compares the obtained rate profiles (labeled with MC-) with the rate profiles of bit-channel cutoff rates at $E_b/N_0 = 1$ and $2$ dB.
As shown in this figure, due to step 1 of algorithm, the resulting rate profile always lies below the cutoff-rate profile at the target SNR.
However, due to step 5, depending on the value of $N$ and $K$, the final rate profile may fall below the cutoff-rate profiles at even smaller SNR values.
For example, because of the removal of $48$ indices from the cutoff-rate profile at $3$ dB, MC-3dB profile falls below the cutoff-rate profile at $2$ dB.
This figure also plots RM-Polar rate profile constructed at $E_b/N_0 = 2.5$ dB which lies on the cutoff-rate profile at $1$ dB.

Fig. \ref{fig: FER_ANV_N256} compares the FER and ANV performances of PAC$(256, 128)$ codes constructed at $E_b/N_0 = 1.5,2.5$ and $3$ dB.
Also plotted is the random-coding union (RCU) bound \cite{polyanskiy2010channel}.
The MC-1.5dB rate profile results in a FER performance similar to the one of RM-Polar rate profile.
As the construction SNR increases, the number of candidate indices in step 2 increases as well, which provides a wider range of indices for step 4 to evaluate.
Hence, increasing the construction SNR can increase the error correction performance of PAC code.
For this reason, MC-2.5dB performs better than MC-1.5dB and
MC-3dB performs better than other both.
The MC-3dB rate profile has approximately $0.5$ dB gain at $\text{FER} = 10^{-3}$ compared to RM-Polar rate profile.

Obviously, increasing the construction SNR may shift the low mean computational decoding region to higher SNR values.
This effect is illustrated by ANV curves in Fig. \ref{fig: FER_ANV_N256}.
Since RM-Polar and MC-1.5dB both lie on cutoff-rate profile at $1$ dB (Fig. \ref{fig: profile_N256}), they both result in a low mean computational complexity beyond $E_b/N_0 = 1$ dB.
On the other hand, MC-2.5dB and MC-3dB lie below cutoff-rate profile at $2$ dB (Fig. \ref{fig: profile_N256}) and thus have low ANV for SNR values beyond $E_b/N_o = 2$ dB.

\begin{figure}[t] 
\centering
	\includegraphics [width = \columnwidth]{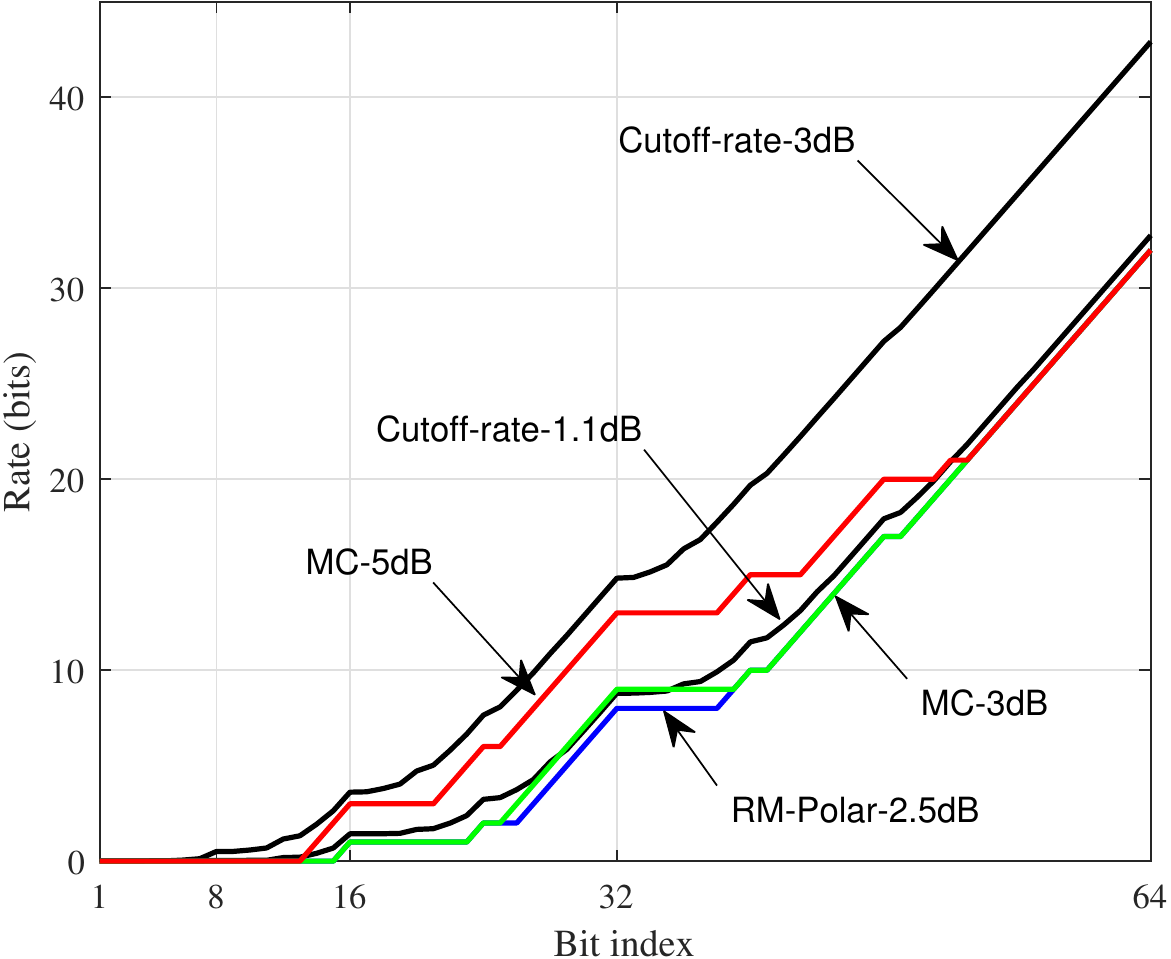}
	\caption{PAC$(64, 32)$ code rate profiles.} 
	\label{fig: profile_N64}
\end{figure}

For PAC$(64, 32)$ code, two different rate profiles are obtained using MC construction method with $\delta = 0.5$ and $E_b/N_0 = 3$ and $5$ dB.
At step 2 of algorithm, $|\mathcal{A}|$ is $42$ and $53$, and to obtain the final rate profile the algorithm runs for $10$ and $21$ iterations, respectively.
Table \ref{tab: rate profile} provides the obtained rate profiles.

Fig. \ref{fig: profile_N64} plots the obtained rate profiles and compares them with the rate profiles of bit-channel cutoff rates at $E_b/N_0 =1.1$ and $3$ dB.
Similar to PAC$(256,128)$, even though MC-5dB is constructed at $E_b/N_0 = 5$, the final rate profile falls below the cutoff-rate profile at $3$ dB because of the removal of 21 indices from initial rate profile.
The RM-Polar and MC-3dB rate profiles lie below the cutoff rate profile at $E_b/N_0 = 1.1$ dB.

\begin{figure}[t] 
\centering
	\includegraphics [width = \columnwidth]{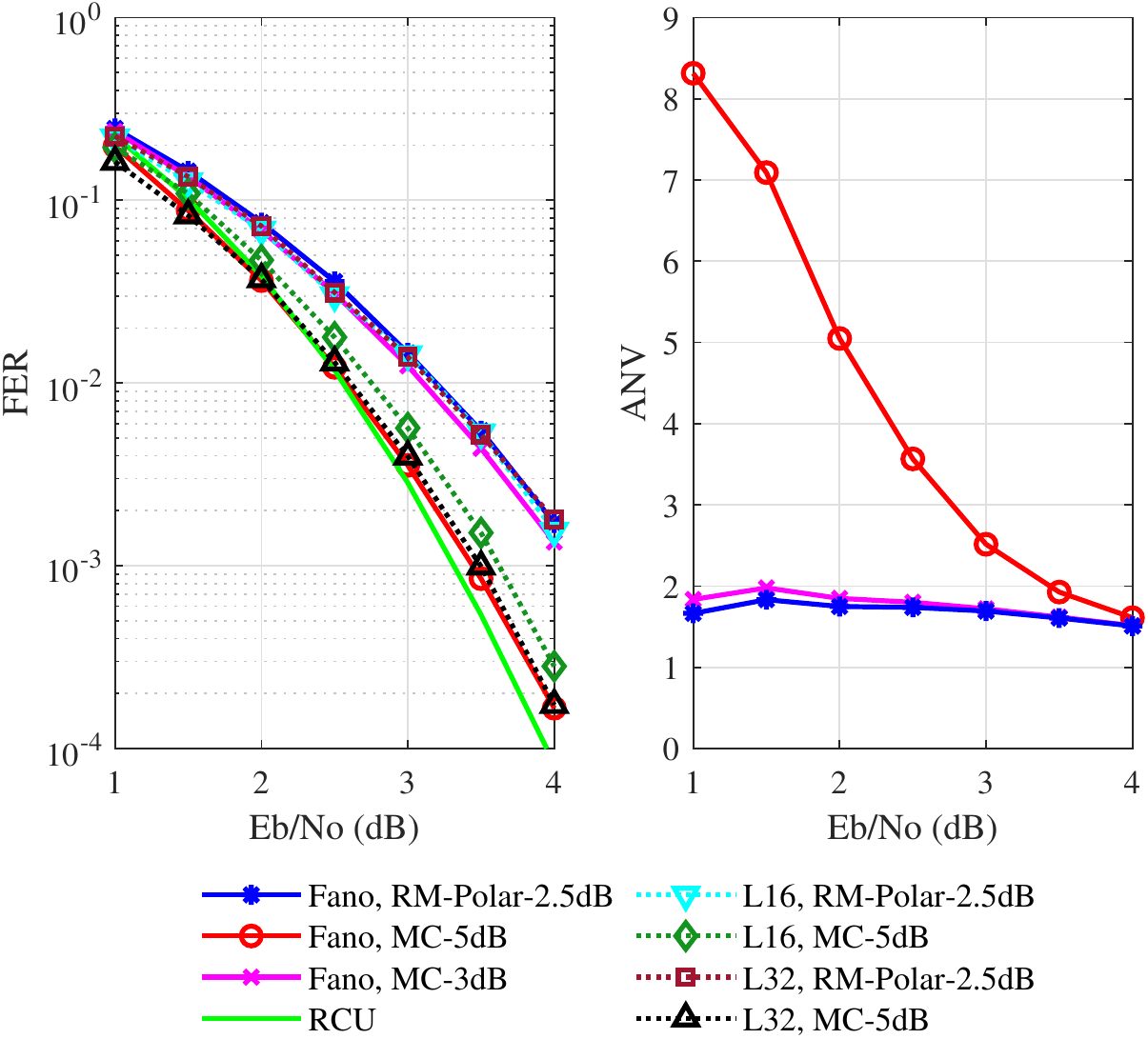}
	\caption{Error-correction performance and complexity of PAC$(64, 32)$ codes.} 
	\label{fig: FER_ANV_N64}
\end{figure}

Fig. \ref{fig: FER_ANV_N64} compares the FER and ANV performances of PAC$(64, 32)$ codes constructed at $E_b/N_0 = 3$ and $5$ dB.
The MC-3dB and RM-Polar rate profile have similar FER and ANV performances.
Since they both lie on cutoff-rate profile at $1.1$ dB (Fig. \ref{fig: profile_N64}), they both result in a low mean computational complexity beyond $E_b/N_0 = 1.1$ dB.
Compared to MC-3dB, the FER performance of MC-5dB is closer to RCU bound but this improvement of FER performance comes at a cost of increased decoding complexity.
The MC-5dB lies below cutoff-rate profile at $3$ dB (Fig. \ref{fig: profile_N64}) and thus has a low ANV for SNR values beyond $E_b/N_o = 3$ dB.

Notice that compared to PAC$(256,128)$, the ANV values of PAC$(64,32)$ are significantly smaller.
For this reason, list decoding of PAC$(64,32)$ code with a small list size is feasible.
The FER performance of list decoding of PAC$(64,32)$ code with MC and RM-Polar rate profiles and list sizes of $16$ and $32$ is plotted in Fig. \ref{fig: FER_ANV_N64}.
To implement the list decoder for PAC codes we used the algorithm introduced in \cite{vardi2021list}.
As the list size increases, the FER performance of list decoding of PAC$(64,32)$ code converges to the one of Fano decoding of PAC$(64,32)$ code for both MC-5dB and RM-Polar rate profiles.
Similar to Fano decoding, the list decoding of PAC$(64,32)$ code has approximately $0.5$ dB gain at $\text{FER} = 10^{-3}$ when using MC-5dB.

\section{Conclusion}\label{sec: conclusion}

In this paper, we proposed a rate-profile construction method for PAC codes, which is capable of maintaining trade-off between the error-correction performance and decoding complexity.
This method aims to improve the FER performance of PAC codes while guaranteeing a low mean sequential decoding complexity for SNR values beyond a target SNR.
Compared to RM-Polar rate profiles, simulation results showed $0.5$ dB coding gain at $\text{FER} = 10^{-3}$ for PAC$(256,128)$ and PAC$(64,32)$ codes constructed with our proposed rate profiles.
Our proposed rate-profile construction method can be applied to any pre-transformed polar code by replacing the generator (Toeplitz) matrix of the CC with a desired matrix.
Also, by replacing the generator matrix of CC by an identity matrix, this method can be applied to conventional polar codes.


\ifCLASSOPTIONcaptionsoff
  \newpage
\fi

\bibliographystyle{IEEEtran}

\end{document}